\documentclass[aps,prd,preprint]{revtex4-1}
\usepackage{graphicx}
\usepackage{amsfonts}
\usepackage{amssymb}
\usepackage{amsmath}

\usepackage{ulem}
\usepackage{color}

\begin{document}
\title{
Determination of Reference Scales for Wilson Gauge Action from
Yang--Mills Gradient Flow
}

\author{Masayuki~Asakawa}
\email{yuki@phys.sci.osaka-u.ac.jp}
\affiliation{Department of Physics, Osaka University, Toyonaka, Osaka 560-0043,
Japan}

\author{Takumi~Iritani}
\email{iritani@yukawa.kyoto-u.ac.jp}
\affiliation{Yukawa Institute for Theoretical Physics, Kyoto
606-8512, Japan}

\author{Masakiyo~Kitazawa}
\email{kitazawa@phys.sci.osaka-u.ac.jp}
\affiliation{Department of Physics, Osaka University, Toyonaka, Osaka 560-0043,
Japan}

\author{Hiroshi~Suzuki}
\email{hsuzuki@phys.kyushu-u.ac.jp}
\affiliation{Department of Physics, Kyushu University, 6-10-1 Hakozaki, Higashi-ku, Fukuoka, 812-8581, Japan}

\collaboration{FlowQCD Collaboration}

\preprint{KYUSHU-HET-152,YITP-15-19}

\date{\today}
\begin{abstract}
  A parametrization of the lattice spacing ($a$) in terms of
  the bare coupling ($\beta$) for the SU(3) Yang--Mills theory with the Wilson
  gauge action is given in a wide range of~$\beta$.
  The Yang--Mills gradient flow with respect to the flow time~$t$
  for the dimensionless observable, $t\frac{d}{dt}t^2\langle E(t)\rangle$, is
  utilized to determine the parametrization. With fine lattice spacings
  ($6.3\le\beta\le7.5$) and large lattice volumes ($N_{\rm s}=64$--$128$), the
  discretization and finite-volume errors are significantly reduced to the same
  level as the statistical error. 
\end{abstract}
\maketitle

\section{Introduction}
\label{sec:intro}
In the asymptotic-free gauge theories such as Yang--Mills theory and QCD,
observables are obtained in the physical unit once the coupling constants of the
theory are fixed at a chosen energy scale. In lattice gauge theory, this
procedure is conveniently accomplished by identifying the energy scale with the
inverse lattice spacing ($a^{-1}$). In actual applications, accurate
determination of the relation between the lattice spacing and the gauge
coupling is crucial not only for obtaining the observables in the physical unit
but also for the continuum extrapolation of the numerical results.

To derive such a relation, the reference scales, e.g.\ the string
tension~\cite{Edwards:1997xf} and the Sommer scale~\cite{Sommer:1993ce}, have
been widely used in the literature. In the determination of these quantities in
the lattice unit, the heavy-quark potential~$V(r)$ for a certain range of the
distance~$r$ needs to be fitted, which inevitably introduces some systematic
errors.

Recently, a novel concept, the Yang-Mills gradient flow, was proposed in~Ref.~\cite{Luscher:2009eq},
and attracts wide attention both analytically and
numerically. It provides us with conceptual and numerical advantages in lattice
simulations~\cite{Luscher:2010iy,Luscher:2011bx,Borsanyi:2012zs,Fodor:2012td,
Fritzsch:2013je,Luscher:2013cpa,Luscher:2013vga,Fodor:2014cpa,Kikuchi:2014rla,
Hasenfratz:2014rna,Ramos:2014kka,Aoki:2014dxa,Monahan:2015lha}.
In particular, the gradient flow can be used to define proper energy-momentum
tensor on the lattice~\cite{Suzuki:2013gza,DelDebbio:2013zaa,Makino:2014taa,
Makino:2014sta,Makino:2014cxa,Suzuki:2015fka}, which opens a new possibility to
study the thermodynamics in lattice field
theories~\cite{Asakawa:2013laa,Kitazawa:2014uxa}.
To  determine the reference scale  with the gradient
flow, a dimensionless observable such as $t^2\langle E(t)\rangle$ (see Eq.~(\ref{eq:E})
for the definition) is measured as a function of the flow
time~$t$ ~\cite{Luscher:2010iy}. Then, $t$ at which the observable takes a specific value is used for
the reference scale. This method does not require the fitting procedure unlike
the previous ones. Moreover, the statistical errors turned out to be
substantially small compared to those of~$V(r)$~\cite{Luscher:2010iy}. A
variant of this method was also proposed in~Ref.~\cite{Borsanyi:2012zs}, where
high-precision scale setting in lattice QCD is attempted.

The purpose of the present paper is to determine the relation between the
lattice spacing ($a$) and the bare coupling ($\beta=6/g_0^2$) in the SU(3)
Yang--Mills theory with the Wilson gauge action over a wide range of~$\beta$
with high accuracy using the Yang--Mills gradient flow. Such a determination is
useful for studying, e.g.\ the precision thermodynamics of the SU(3) Yang--Mills
theory~\cite{Asakawa:2013laa,Kitazawa:2014uxa}. In the previous studies with
the Wilson gauge action~\cite{Edwards:1997xf,Guagnelli:1998ud,Necco:2001xg},
the range of~$\beta$ covered was $5.6\le\beta\le6.92$. On the other hand, in
the present study, by exploiting the benefit of the gradient flow together with
the recent advance in computational power, we go into a weaker coupling region,
$\beta=6.3$--$7.5$. Moreover, to suppress the finite volume effect for the two
finest lattices ($\beta=7.4$ and $7.5$), the lattice volume~$128^4$ is taken for
those cases.

Following the procedures proposed
in~Refs.~\cite{Luscher:2010iy,Borsanyi:2012zs}, we consider the dimensionless
observables $t^2\langle E(t) \rangle$ and~$td(t^2\langle E\rangle)/dt$ and fix
their values to be~$X$ to determine the reference scale. We vary $X$ in order
to suppress both the lattice spacing and finite volume effects below the
magnitude of the statistical error. This enables us to derive accurate relation
between $a$ and~$\beta$ for the wide range of lattice spacing with an accuracy
of less than~$0.5\%$.

This paper is organized as follows. In Sec.~\ref{sec:GF}, we introduce the
gradient flow and define our reference scales. We then present our numerical
results of the dimensionless observables in Sec.~\ref{sec:num}. After
describing our numerical setup, detailed analyses on the discretization and
finite volume effects are performed. The parametrization of the lattice spacing
in terms of~$\beta$ and comparisons with previous studies without the 
gradient flow are also presented in
this section. 
In Sec.~\ref{sec:summary}, we give a brief summary.
 
\section{Gradient flow and reference scales}
\label{sec:GF}
The gradient flow is a
continuous transformation of fields; for gauge fields, it is defined by the
differential equation~\cite{Luscher:2009eq,Luscher:2010iy},
\begin{equation}
   \frac{d A_\mu}{dt}
   =-g_0^2\frac{\partial S_{\mathrm{YM}}(t)}{\partial A_\mu}
   =D_\nu G_{\nu\mu},
\label{eq:GF}
\end{equation}
with the Yang--Mills action $S_{\mathrm{YM}}(t)$ defined by $A_\mu(t)$. Color
indices are suppressed for simplicity. The $A_\mu(0)$ is identified with the
standard gauge field defined in four dimensional space-time. The flow time~$t$,
having a dimension of inverse mass-squared, controls the flow into the extra
dimension. The gauge field is transformed along the steepest descent direction
of~$S_{\mathrm{YM}}(t)$ as $t$ increases. In the tree level, Eq.~(\ref{eq:GF}) is
rewritten as 
\begin{equation}
   \frac{d A_\mu}{dt}
   =\partial_\nu\partial_\nu A_\mu+\text{(gauge dependent terms)},
\label{eq:diffusion}
\end{equation}
which is essentially a diffusion equation. For positive~$t$, therefore, the
gradient flow acts as the cooling of the gauge field with the smearing
radius~$\sqrt{8t}$. In~Ref.~\cite{Luscher:2011bx}, it is rigorously proved that
all composite operators composed of~$A_\mu(t)$ take finite values for~$t>0$.
This property ensures that observables at~$t>0$ are automatically renormalized.

One of the composite operators whose $t$ dependence is extensively studied is
\begin{equation}
   E(t)=\frac{1}{4}G_{\mu\nu}^a(t)G_{\mu\nu}^a(t),
\label{eq:E}
\end{equation}
where $G_{\mu\nu}^a(t)$ is the ``field strength'' composed of~$A_\mu(t)$. The
$t$ dependence of the vacuum expectation value of~Eq.~(\ref{eq:E}) for
small~$t$ is obtained perturbatively up to next-to-leading order
as~\cite{Luscher:2010iy}
\begin{equation}
   t^2\langle E(t)\rangle
   =\frac{3(N_c^2-1)g(q)^2}{128\pi^2}
   \left[1+\frac{k_1}{4\pi}g(q)^2+O(g(q)^4)\right],
\label{eq:t2Epert}
\end{equation}
with~$k_1=N_c(11\gamma_E/3+52/9-3\ln3)/(4\pi)$ in the $\overline{\mathrm{MS}}$
scheme and $N_c$ being the number of color. The running coupling~$g(q)$ is
defined at the scale of the smearing radius, $q=1/\sqrt{8t}$.
Equation~(\ref{eq:t2Epert}) implies that the dimensionless quantity
$t^2\langle E(t)\rangle$ is 
 an increasing function of~$t$ for small flow time 
  corresponding to the perturbative regime. As
shown numerically in~Ref.~\cite{Luscher:2010iy} on the lattice, $t^2\langle E(t)\rangle$
 is also a monotonically increasing function for large $t$
  corresponding to the  non-perturbative regime. Therefore, the
value of~$t$ at which $t^2\langle E(t)\rangle$ takes a specific value~$X$,
i.e., the solution of the equation
\begin{equation}
   \left.t^2\langle E(t)\rangle\right|_{t=t_{_X}}=X,
\label{eq:t_x}
\end{equation}
is expected to be a unique dimensionful quantity, which can be used as a reference scale to
introduce physical unit in lattice gauge theory. In~Ref.~\cite{Luscher:2010iy},
$t_{_{X}=0.3}$ (sometimes called~$t_0$) is used as the reference scale.
In~Ref.~\cite{Borsanyi:2012zs}, a quantity~$w_{_X}$ defined by 
\begin{equation}
   \left.t\frac{d}{dt}t^2\langle E(t)\rangle\right|_{t=w_{_X}^2}=X,
\label{eq:w_x}
\end{equation}
is proposed as an alternative reference scale. In~Ref.~\cite{Borsanyi:2012zs},
a reference scale~$w_{_{X}=0.3}$ (sometimes called~$w_0$) is employed to set the
scale and it was found that the discretization error of~$w_{0.3}$ is
suppressed more than that of~$t_{0.3}$ in full QCD~\cite{Borsanyi:2012zs}.

In the present study we consider more general reference scales, $t_{_X}$
and~$w_{_X}$ with $X=0.2$, $0.3$ and~$0.4$.
Larger $X$ is preferable to suppress the lattice discretization
error~\cite{Fodor:2014cpa}, while the smearing radius~$\sqrt{8t}$ would
eventually hit the lattice boundary for too large~$X$. We note that the
numerical cost increases as $X$ increases, since more time-steps are required
for solving the differential equation Eq.~(\ref{eq:GF}).
We use $w_{0.4}$ and~$w_{0.2}$ for the reference scales and introduce a
new parametrization of the lattice spacing~$a$ in terms of the bare
coupling~$\beta=6/g_0^2$ by a hybrid use of these reference scales.

\section{Numerical results}
\label{sec:num}
\subsection{Simulation setup}
\label{sec:setup}
We perform numerical analyses of  the SU(3) Yang--Mills
theory with the Wilson plaquette action in the present study.
Gauge configurations are generated by
a combination of one heatbath and five overrelaxation updates; we refer this
set for updates as one Monte Carlo update. We perform $20,000$ Monte Carlo
updates for the thermalization from the cold start, and analyze configurations
separated by $1,000$ updates after the thermalization. We take the periodic
boundary condition with the lattice size~$N_{\rm s}^4$. The values of~$\beta=6/g_0^2$,
$N_{\rm s}$ and the number of configurations $N_{\mathrm{conf}}$ are summarized
in~Table~\ref{table:param}. For $\beta=7.0$, $7.2$ and~$7.4$, we take two
different values of~$N_{\rm s}$ to investigate the finite volume effect. In our main
analysis, we use the data with~$N_{\rm s}=64$ for~$\beta=6.3$--$6.9$, $N_{\rm s}=96$
for~$\beta=7.0$--$7.2$ and $N_{\rm s}=128$ for~$\beta=7.4$--$7.5$. These data sets
are indicated by~$*$ in the last column of the Table. As we will show
in~Sec.~\ref{sec:V}, the finite volume effect is well suppressed for these
choices.

Autocorrelation between different configurations is analyzed by the dependence
of the jackknife statistical errors against the bin-size, $N_{\mathrm{bin}}$, 
together with the autocorrelation function.
For~$N_{\rm s}=64$, we found no $N_{\mathrm{bin}}$ dependence, so that different
configurations are uncorrelated. For $N_{\rm s}=96$ and~$128$, the statistical error
increases up to about~$20\%$ as $N_{\mathrm{bin}}$ increases
in~$N_{\mathrm{bin}}\le5$. Then, we make the jackknife error estimate
with~$N_{\mathrm{bin}}=2$ in these cases.
The autocorrelation function shows that
the autocorrelation is not visible within statistics already 
with one separation of $1,000$ updates, 
which is consistent with the analysis of the bin-size dependence.

The autocorrelation of the topological charge
is known to become longer as the lattice spacing becomes
finer due to the critical slowing down \cite{Luscher:2009eq}.
Our analyses with separations of $1,000$ updates may
suffer from this problem and therefore should be understood with
reservation. New measurements of the topological charge with
careful choice of the flow-time step size 
are left for our future work. 

The lattice discretization of the flow equation Eq.~(\ref{eq:GF}) and that of
the observable~$E$ are not unique~\cite{Luscher:2010iy}. We use the Wilson
gauge action~$S_{\mathrm{YM}}$ for the flow equation
in~Eq.~(\ref{eq:GF})~\cite{Luscher:2010iy}. To construct the operator~$E$, we
use the clover-type representation of~$G_{\mu\nu}^a$ 
unless otherwise stated.
Other choices of~$S_{\mathrm{YM}}$ and~$E$ may reduce the discretization effect
further~\cite{Fodor:2014cpa}.
To solve the differential equation Eq.~(\ref{eq:GF}) numerically,
we use the second-order Runge--Kutta (RK) method. 
To estimate the numerical error of the RK method, 
we have solved the RK with two different integration step sizes,
one of which is twice coarser than the other, for several configurations.
By comparing these results we have checked that the
numerical error of the RK method is within two orders of 
magnitude smaller than the statistical errors.
(This procedure may be improved by the method 
suggested and tested in Refs.~\cite{Fritzsch:2013je}.)

\begin{figure}[]
\centering
\includegraphics[width=0.47\textwidth,clip]{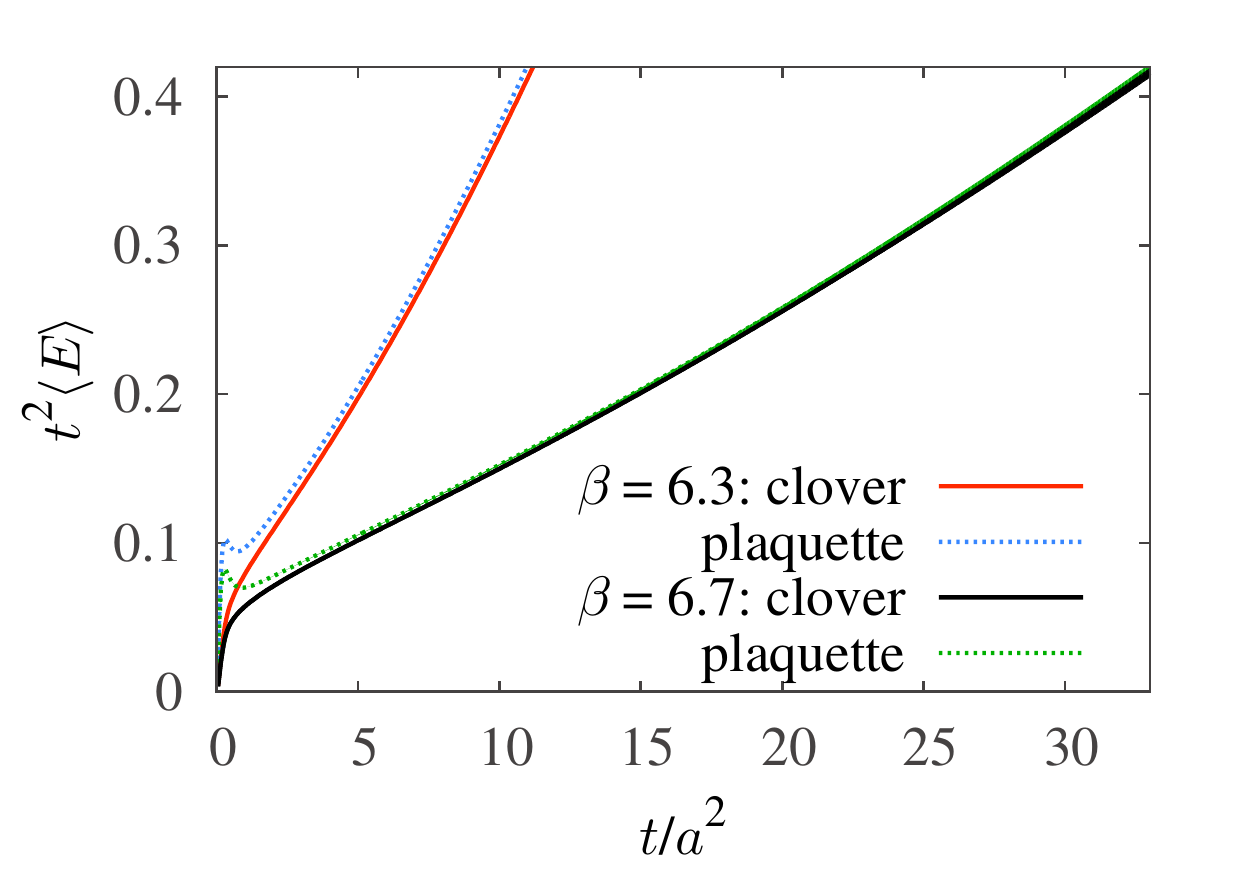}
\includegraphics[width=0.47\textwidth,clip]{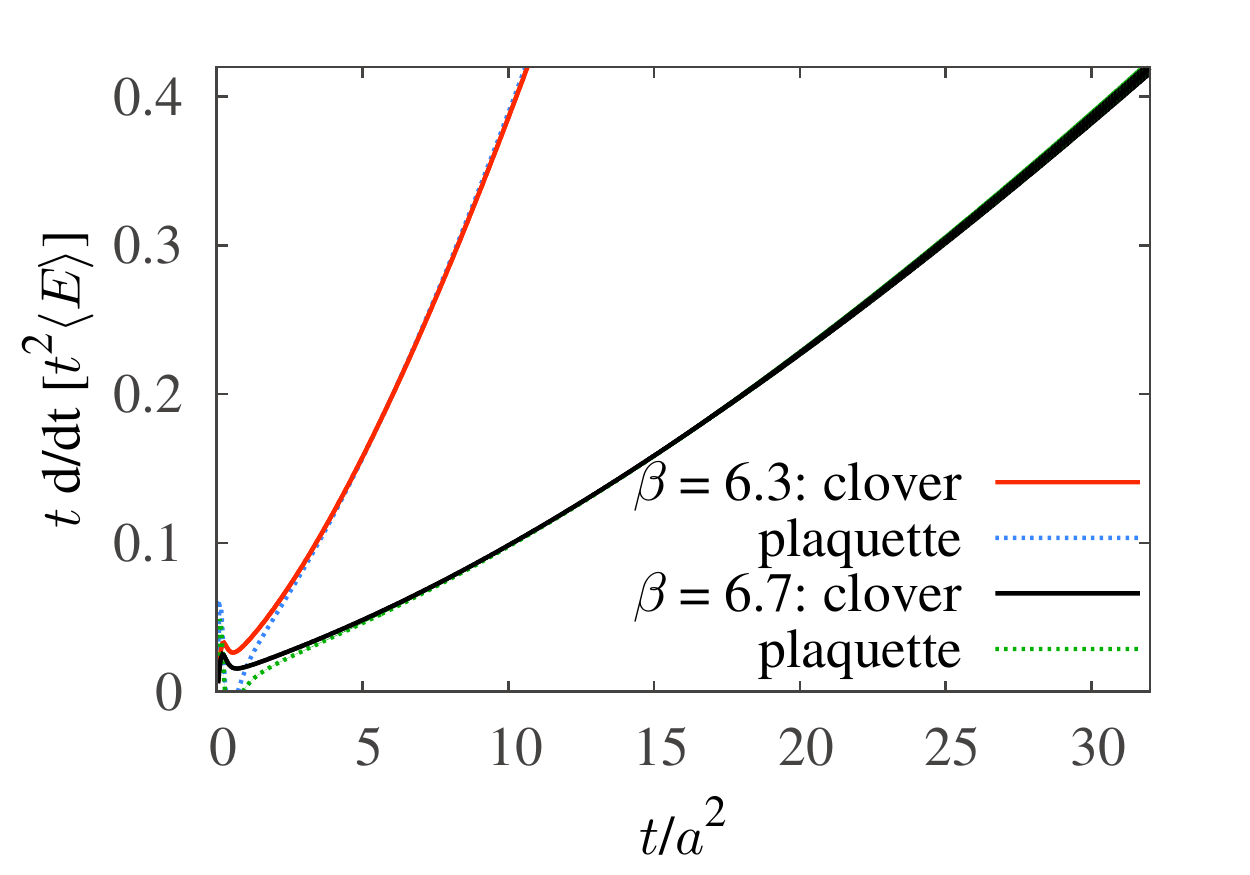}
\caption{$t^2\langle E(t)\rangle$ (the left panel)
and~$t\frac{d}{dt}t^2\langle E(t)\rangle$ (the right panel) as functions
of~$t/a^2$ for~$\beta=6.3$ and~$6.7$. Together with the result obtained with
the clover-type operator for~$E$, the one with~$E$ defined from the plaquette 
is also presented for each parameter. The statistical error is smaller than the
width of the lines.}
\label{fig:core-fig}
\end{figure}

To illustrate the behaviors of~$t^2\langle E(t)\rangle$
and~$t\frac{d}{dt}t^2\langle E(t)\rangle$, we show these quantities with
statistical errors as a function of~$t/a^2$ in~Fig.~\ref{fig:core-fig} for
$\beta=6.3$ and~$6.7$.
To check the discretization effect on the operator, we compare $t^2\langle E(t)\rangle$
and $t\frac{d}{dt}t^2\langle E(t)\rangle$ defined from the clover-type
representation with those defined from~$E(t)=2(1-P(t))$ using the average
plaquette~$P(t)$~\cite{Luscher:2010iy}. The figure shows that the difference
between the two definitions is suppressed for large~$t$. In particular, the
difference is more suppressed in~$t\frac{d}{dt}t^2\langle E(t)\rangle$ than
that in~$t^2\langle E(t)\rangle$, i.e., the discretization effect in the former
is smaller than the latter. The figure also shows that both quantities increase
monotonically as~$t/a^2$ increases except for the small~$t/a^2$ region where the
lattice distortion effect is sizable.
Furthermore, they show approximately linear behavior within the range
$0.2\le X\le0.4$. In~Table~\ref{table:param}, the values of~$w_{_X}(a)$
and~$t_{_X}(a)$ obtained from~Eqs.~(\ref{eq:t_x}) and~(\ref{eq:w_x})
with~$X=0.2$, $0.3$ and~$0.4$ are given for each set of parameters. For
simplicity, we will use the notations, $w_{_X}$ and $t_{_X}$, even for non-zero
values of $a$.
Note that the variances of $w_{_X}$ and~$t_{_X}$ are approximately 
proportional to the inverse of physical volume of the lattice, 
while their dependence on $a$ should be small because gauge invariant 
operators at positive flow time is automatically renormalized 
\cite{Luscher:2010iy}.
The statistical error shown in Table~\ref{table:param} indeed 
shows such dependences on the spatial volume and lattice spacing.

\begin{table}[]
\centering
\caption{Simulation parameters $\beta$ and~$N_{\rm s}$, and number of configurations,
$N_{\mathrm{conf}}$. The result for the values of~$w_{_X}/a$
and~$\sqrt{t_{_X}}/a$ are also presented.}
\label{table:param}
\begin{tabular}{rrr|ccc|ccc|r|c}
    \hline \hline
    $\beta$ & $N_{\rm s}$ & $N_\mathrm{conf}$ & $w_{0.4}/a$ 
    & $w_{0.3}/a$ & $w_{0.2}/a$ & $\sqrt{t_{0.4}}/a$ & $\sqrt{t_{0.3}}/a$ & $\sqrt{t_{0.2}}/a$ & $N_{\rm s}a/w_{0.2}$ &   \\
    \hline 
    6.3 & 64 & 30 &   3.208(7)              & 2.877(5)  & 2.460(4)   &  3.269(4)  & 2.835(3)   & 2.254(2)  & 26.02 & * \\
    6.4 & 64 &100 &   3.697(5)              & 3.317(4)  & 2.837(3)   &  3.765(3)  & 3.263(3)   & 2.590(2)  & 22.56 & * \\
    6.5 & 64 & 49 &   4.231(10)             & 3.797(8)  & 3.249(7)   &  4.310(8)  & 3.736(6)   & 2.965(4)  & 19.70 & * \\
    6.6 & 64 &100 &   4.857(11)             & 4.356(9)  & 3.725(7)   &  4.938(8)  & 4.277(6)   & 3.390(4)  & 17.18 & * \\
    6.7 & 64 & 30 &   5.558(27)             & 4.980(23) & 4.252(17)  &  5.638(20) & 4.878(16)  & 3.863(10) & 15.05 & * \\
    6.8 & 64 &100 &   6.300(20)             & 5.652(17) & 4.833(13)  &  6.406(16) & 5.548(12)  & 4.395(08) & 13.24 & * \\
    6.9 & 64 & 30 &   7.165(62)             & 6.431(52) & 5.503(40)  &  7.289(47) & 6.313(36)  & 5.000(23) & 11.63 & * \\
    7.0 & 64 &209 &   8.177(34)             & 7.322(28) & 6.250(21)  &  8.287(25) & 7.168(19)  & 5.674(12) & 10.24 &  \\
    7.0 & 96 & 60 &   8.137(21)             & 7.297(18) & 6.236(13)  &  8.264(16) & 7.154(12)  & 5.665(08) & 15.39 & * \\
    7.2 & 64 &204 &  10.843(102)            & 9.646(80) & 8.176(58)  & 10.833(69) & 9.326(51)  & 7.343(31) &  7.83 &   \\
    7.2 & 96 & 53 &  10.428(78)             & 9.348(66) & 7.984(52)  & 10.586(62) & 9.162(48)  & 7.256(31) & 12.02 & * \\
    7.4 & 96 & 52 &                         &           & 10.426(106)&            & 11.927(98) & 9.415(64) &  9.21 &  \\
    7.4 &128 & 40 & & 12.084(61) & 10.306(42) &            & 11.808(40)& 9.337(28)  & 12.42 & * \\
    7.5 &128 & 60 & & & 11.706(72) & &        &10.601(42) & 10.94 & * \\
    \hline
\end{tabular}
\end{table}

\subsection{Estimation of $w_{_X}$ for large $X$ in high $\beta$ region}
\label{sec:a}
\begin{figure}[]
\centering
\includegraphics[width=0.47\textwidth,clip]{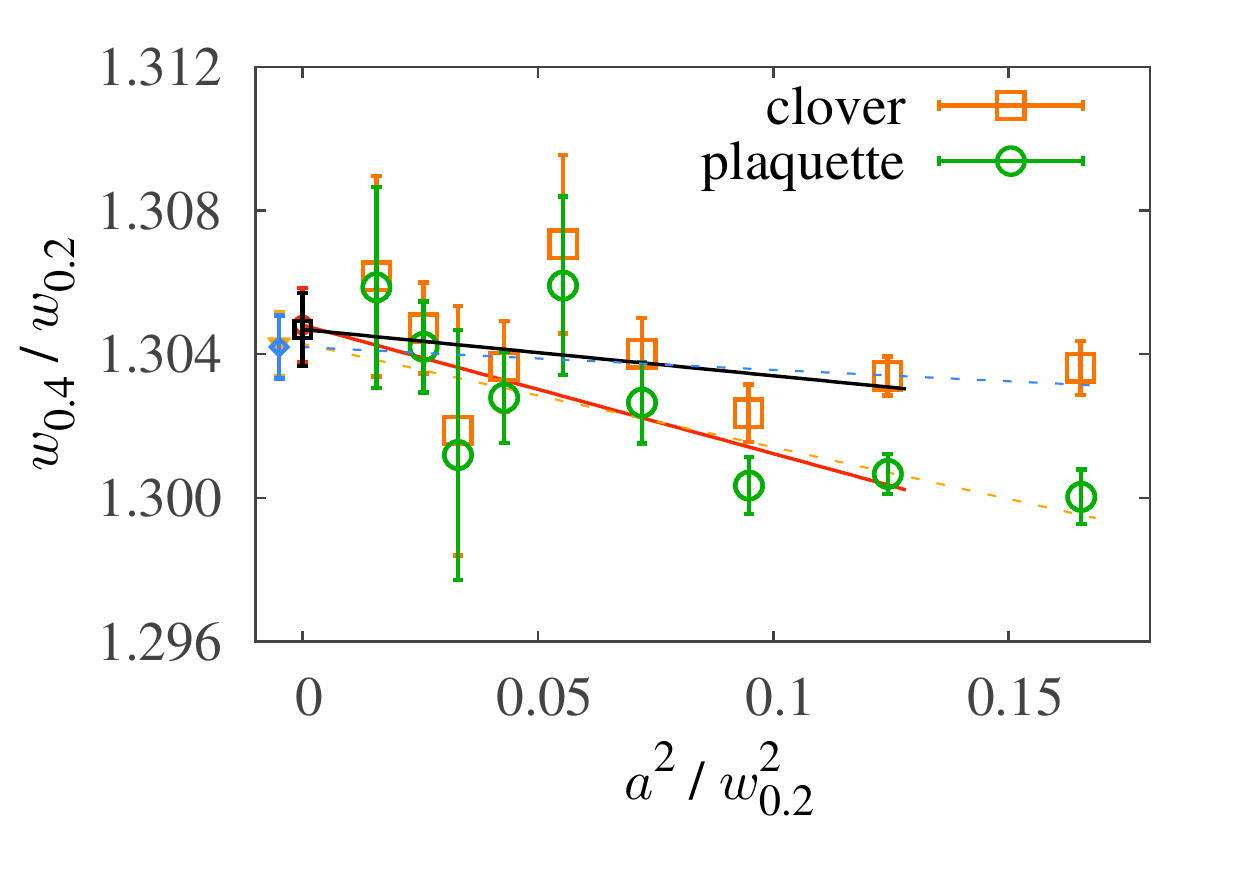}
\includegraphics[width=0.47\textwidth,clip]{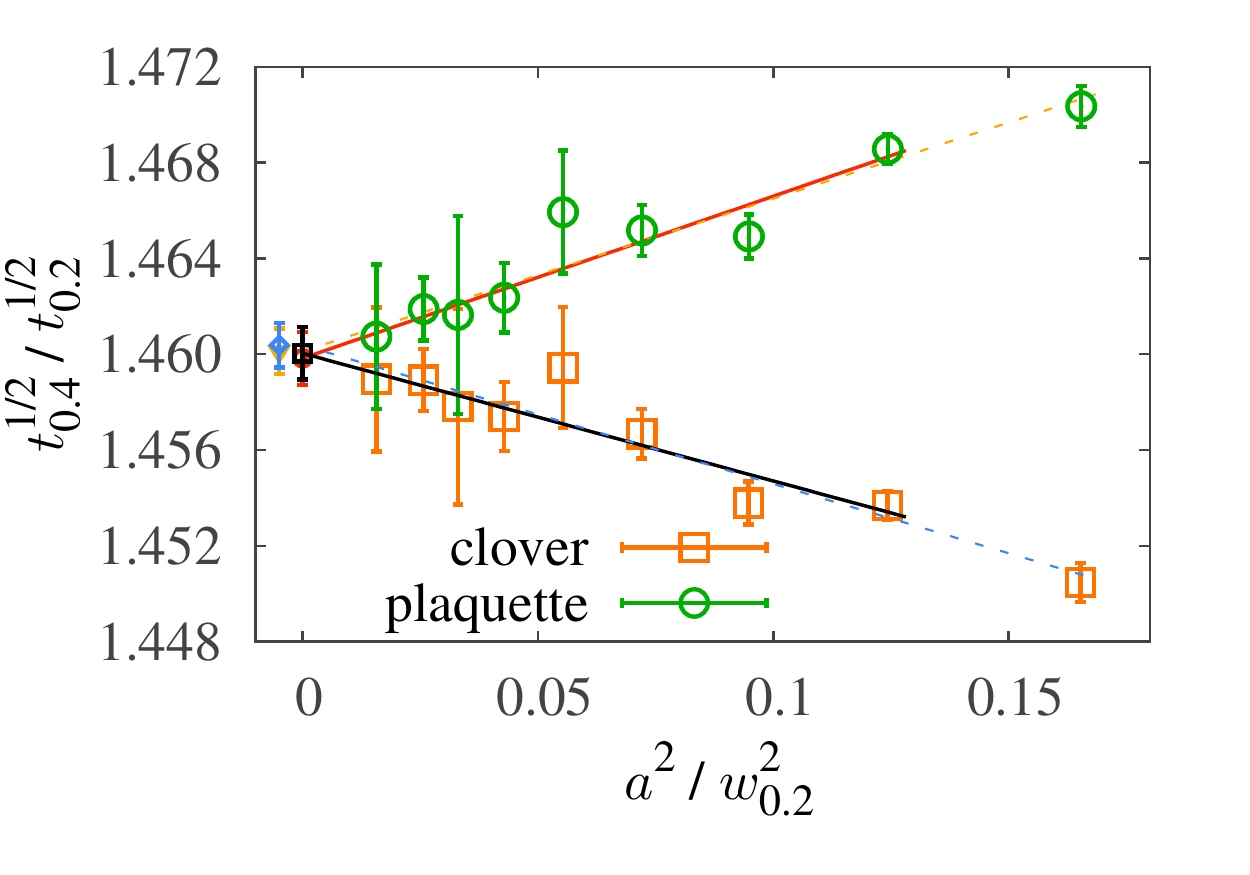}
\caption{Ratios $w_{0.4}/w_{0.2}$ and~$\sqrt{t_{0.4}/t_{0.2}}$ as a function of
the lattice spacing. 
Squares represent the results with the clover-type representation of $E$,
while circles show those with plaquette one.
Solid and dashed lines represent $8$-point and $9$-point
linear fits and the extrapolated values to the continuum limit are shown
at~$a^2/w_{0.2}^2=0$ and~$-0.005$, respectively.}
\label{fig:wt}
\end{figure}

As pointed out in~Ref.~\cite{Borsanyi:2012zs}, the discretization error
of~$w_{0.3}$ is smaller than $t_{0.3}$, so that we employ $w_{_X}$ as the key
reference scale in this paper. 
Note that the lattice artifact is expected to be smaller for larger~$X$.
%

In our simulations, we estimate $w_{0.4}/a$ for $\beta=7.4$, $7.5$
using the data at small flow time ($w_{0.2}/a$) for these $\beta$ 
and the extrapolation of the ratio $w_{0.4}/w_{0.2}$ to the continuum limit.
We plot $9$ data points for $w_{0.4}/w_{0.2}$ in 
the interval $6.3\le\beta\le7.2$ as a
function of~$a^2/w_{0.2}^2$ in the left panel of~Fig.~\ref{fig:wt}.
In the figure, the results are shown for clover- and plaquette-type
representations of $E$.
The error bars of the data points are estimated by the jackknife
method for the ratios and not for the individuals. 
To take the continuum
limit, we perform a linear fit with all $9$~points and that with $8$~points by
removing the data for the coarsest lattice. The same procedure for the
continuum extrapolations will be adopted throughout this paper. 
The values in the continuum limit obtained from~Fig.~\ref{fig:wt} 
with the clover representation
are
$(w_{0.4}/w_{0.2})_{a\to0}=1.3042(9)$ and~$1.3047(10)$. 
The continuum extrapolation with the plaquette representation 
agrees well with this result.
Similar extrapolation
for~$t_{_X}$ shown in the right panel of~Fig.~\ref{fig:wt} leads
to~$(\sqrt{t_{0.4}/t_{0.2}})_{a\to0}=1.4604(9)$ and~$1.4600(11)$.

 The left panel of Fig.~\ref{fig:wt} indicates that
 the lattice discretization error for the ratio $w_{0.4}/w_{0.2}$ is small. 
 With this fact in mind, we estimate the values of~$w_{0.4}/a$ at~$\beta=7.4$
and~$7.5$ using the numerical results of~$w_{0.2}/a$ and the linear fit shown
in~Fig.~\ref{fig:wt} (left). The results are $w_{0.4}/a=13.445(55)(5)$
at~$\beta=7.4$ and $w_{0.4}/a=15.272(95)(6)$ at~$\beta=7.5$. The first error in
the parenthesis is from the statistical error of~$w_{0.2}/a$ and the fit
parameters, while the second error is the systematic error obtained from the
$9$-point linear fit. The latter is more than one order of magnitude smaller
than the former.

\subsection{Effect of the finite volume}
\label{sec:V}
In order to investigate the finite volume effect, we have performed the
numerical analyses for two different values of~$N_{\rm s}$ for~$\beta=7.0$, $7.2$
and~$7.4$ as shown in~Table~\ref{table:param}, where the spatial sizes in
physical unit normalized by~$w_{0.2}$, $L/w_{0.2}=N_{\rm s}a/w_{0.2}$, are shown for
each set of configurations.

The comparison of the results with different $N_{\rm s}$ for~$\beta=7.0$ and~$7.4$
in~Table~\ref{table:param} shows that the values of~$w_{_X}/a$
and~$\sqrt{t_{_X}}/a$ for different~$N_{\rm s}$ agree within the statistics. On the
other hand, the results for~$\beta=7.2$ have statistically significant $N_{\rm s}$
dependence between $N_{\rm s}=64$ and~$96$. These results suggest that the finite
volume effect modifies the numerical results for~$L/w_{0.2}=7.83$, while the
effect is not visible for~$L/w_{0.2}>9.21$ in the present statistics. This is
the reason why we use the data sets with~$*$ in the last column
of~Table~\ref{table:param}.
 
\subsection{Parametrization by the bare coupling $\beta$}
\label{sec:parametrization}
For various practical applications, it is convenient to introduce a
parametrization of the ratio~$w_{0.4}/a$ in terms of~$\beta$. We have carried
out such parametrization using four types of fitting functions (polynomial
type, one-loop type, Pad\'e type, two-loop type) as summarized
in~Appendix~\ref{sec:fit}: All these fitting functions can reproduce the
numerical results in~Table.~\ref{table:param} well with three or four
parameters. Among them, the three parameter fit motivated by the one-loop
perturbation theory provides a reasonable result
($\chi^2/{\mathrm{dof}}=0.917$) for $11$ data points in~$6.3\le\beta\le7.5$
without over fitting:
\begin{equation}
   \frac{w_{0.4}}{a}
   =\exp\left(\frac{4\pi^2}{33}\beta
   -8.6853
   +\frac{37.422}{\beta}
   -\frac{143.84}{\beta^2}
   \right)
   [1\pm0.004(\mathrm{stat.})\pm 0.007(\mathrm{sys.})] 
\label{eq:fitfinal}
\end{equation}

\begin{figure}
\centering
\includegraphics[width=0.47\textwidth,clip]{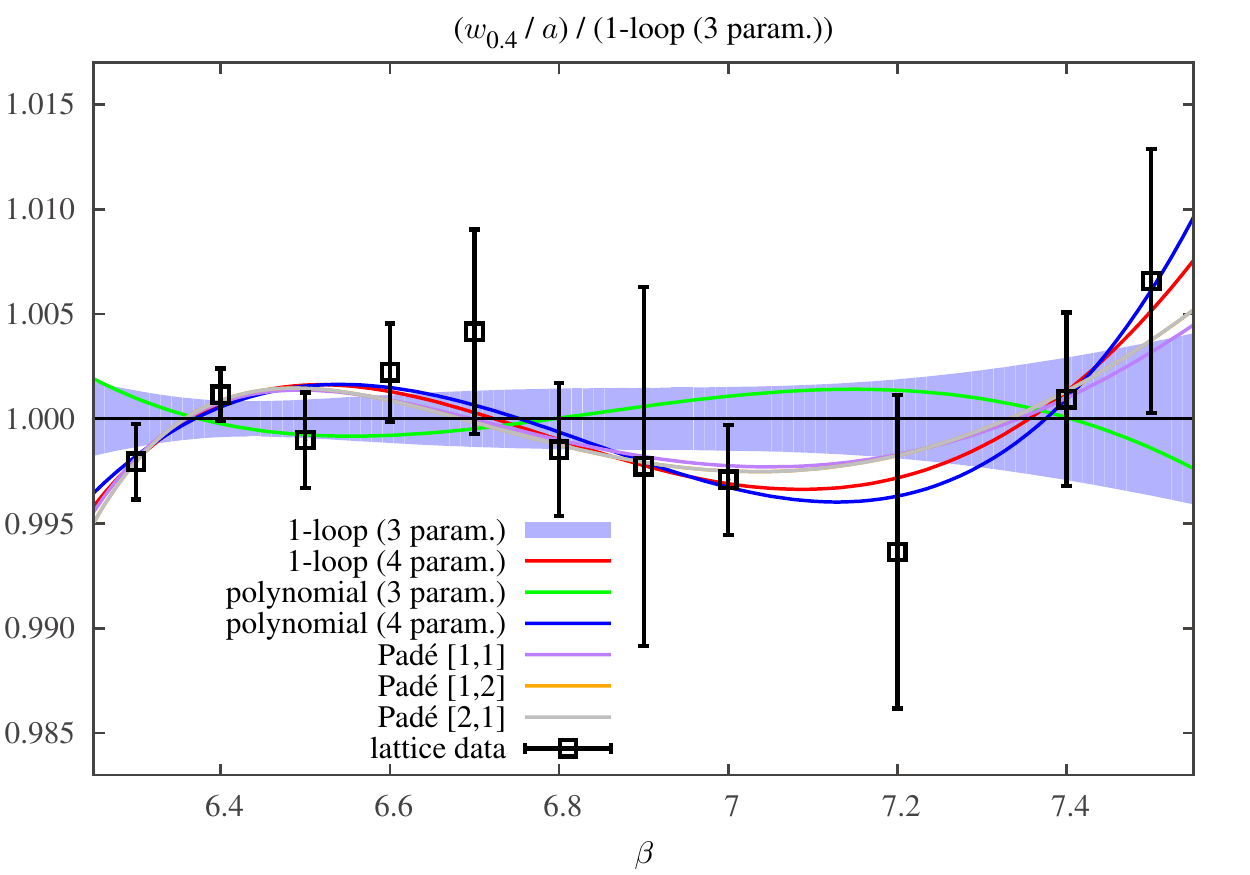}
\includegraphics[width=0.47\textwidth,clip]{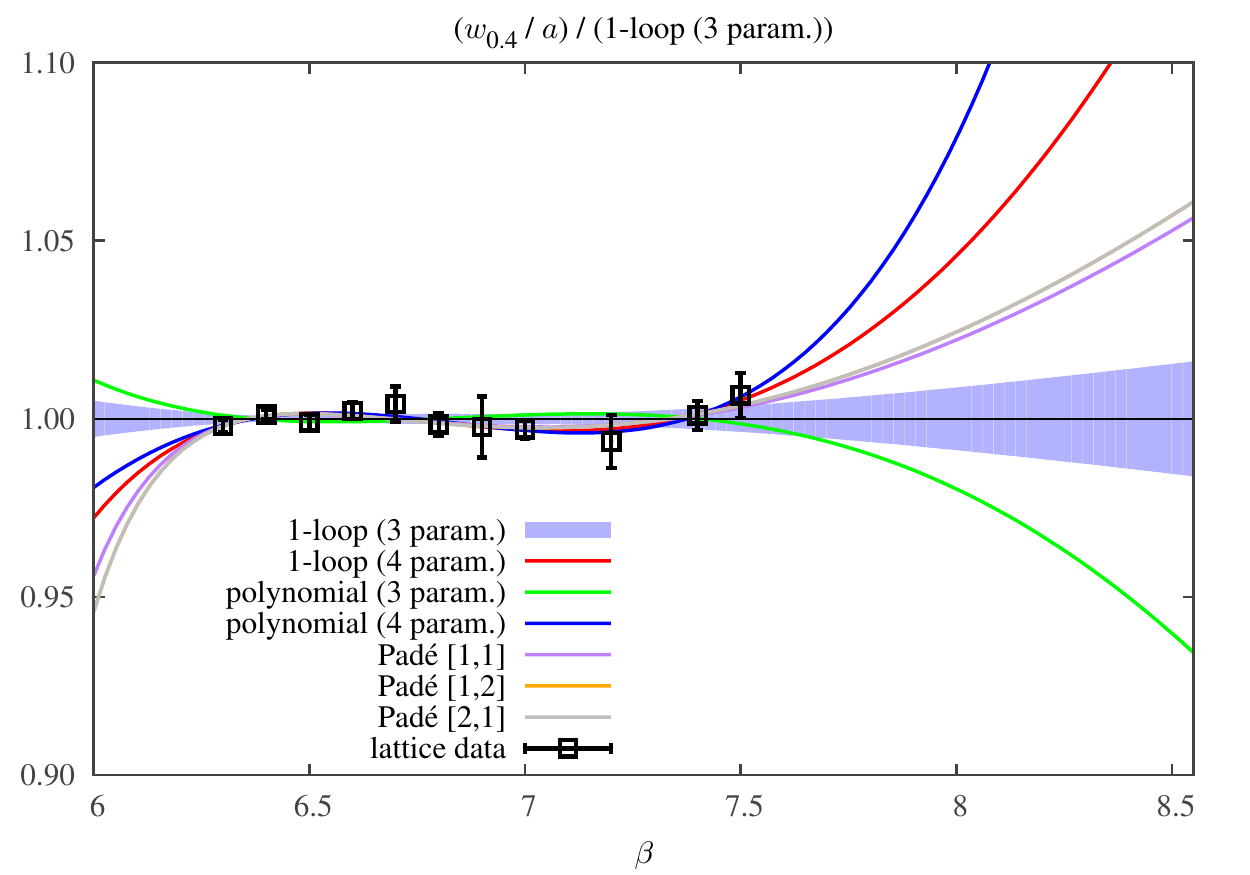}
\caption{(Left) Result of the three parameter fit of~$w_{0.4}/a$ given
in~Eq.~(\ref{eq:fitfinal}). Squares are the data in~Table~\ref{table:param}
normalized by~Eq.~(\ref{eq:fitfinal}). Shaded band indicates the uncertainty
originating from the errors of the fitted coefficients. Results with several
other fitting functions normalized by~Eq.~(\ref{eq:fitfinal}) are also plotted
as well; see, Appendix~\ref{sec:fit}. (Right) The same result is plotted for a
wide range of~$\beta$.}
\label{fig:fit}
\end{figure}

In the left panel of~Fig.~\ref{fig:fit}, we show the numerical results
of~Table~\ref{table:param} normalized by the fitting function
Eq.~(\ref{eq:fitfinal}). The shaded band in~Fig.~\ref{fig:fit} is the error
associated with the fitting parameters in~Eq.~(\ref{eq:fitfinal}). The results
of some other fitting functions in~Appendix~\ref{sec:fit} normalized
by~Eq.~(\ref{eq:fitfinal}) are also plotted in Fig.~\ref{fig:fit}. They agree
with each other within~$0.5\%$ in the range, $6.3\le\beta\le7.5$. In the right
panel of~Fig.~\ref{fig:fit}, the fitting functions are plotted in the
region beyond the present $\beta$. Although the difference among the curves
grows as $\beta$ becomes large, the deviation is still within~$7\%$ even
at~$\beta=8.0$.

\subsection{Continuum extrapolation of $w_{_X}$ and $t_{_X}$}
\label{sec:our_ratio}
\begin{figure}[]
\centering
\includegraphics[width=0.47\textwidth,clip]{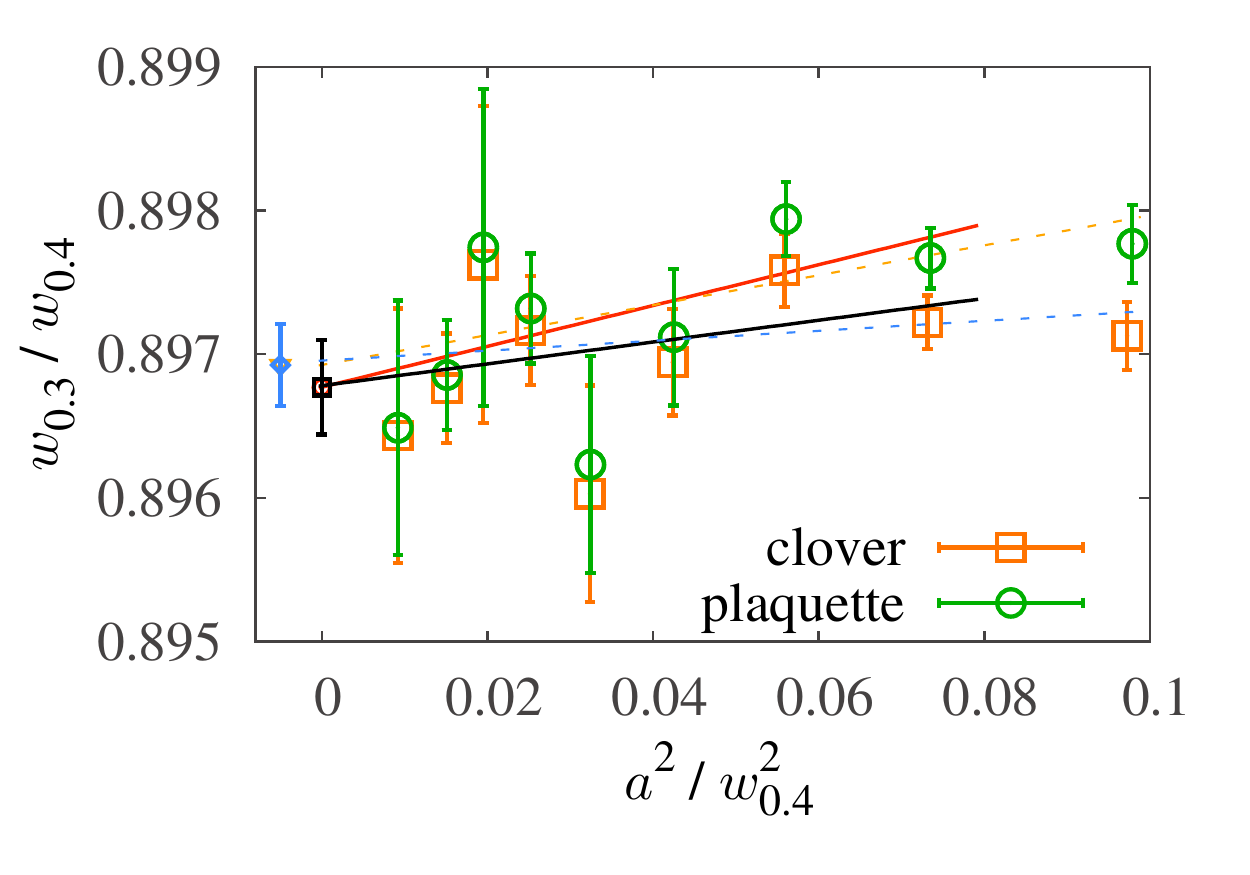}
\includegraphics[width=0.47\textwidth,clip]{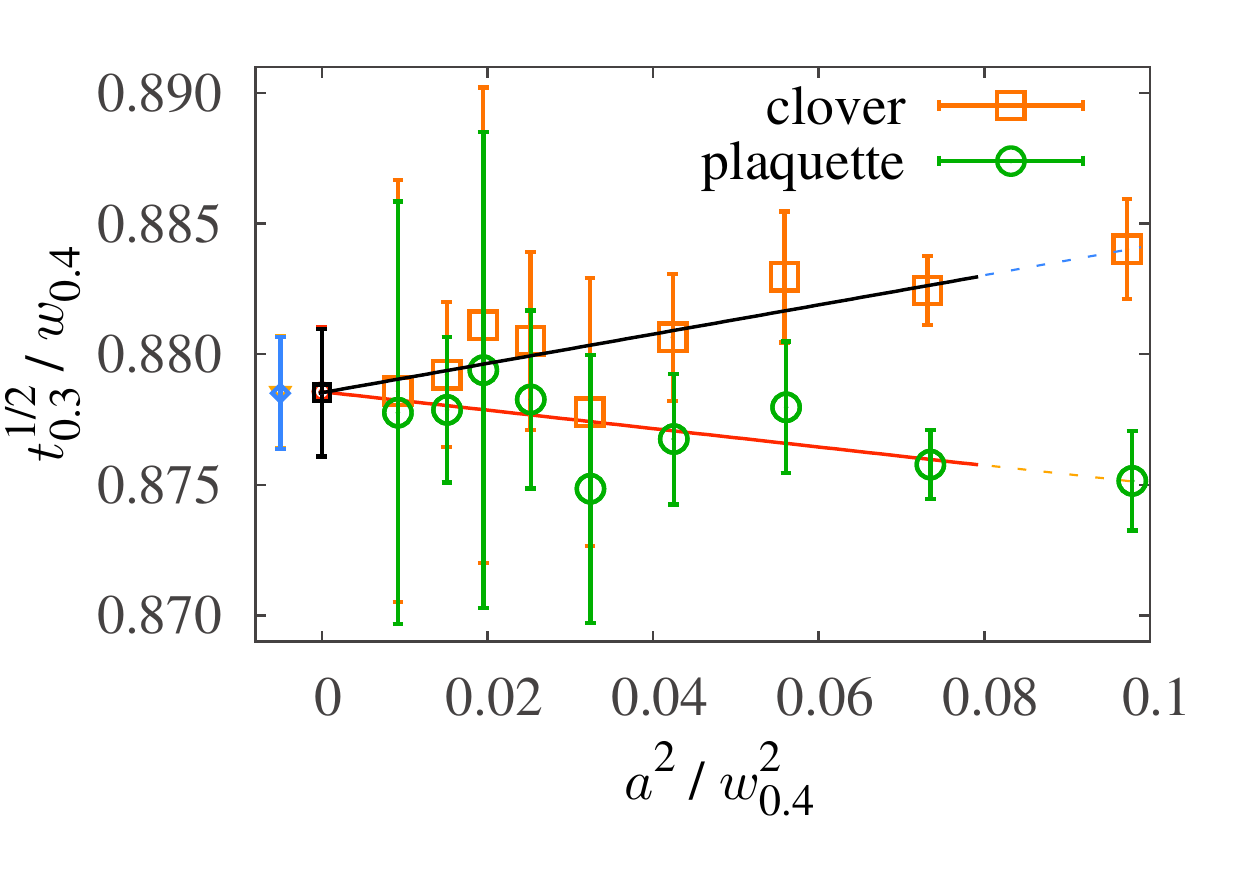}
\caption{Continuum extrapolations of the ratios $\sqrt{t_{0.3}}/w_{0.4}$
and~$w_{0.3}/w_{0.4}$.
Squares represent the results with the clover-type representation of $E$,
and circles are those with plaquette one.}
\label{fig:wt2}
\end{figure}

We extract the continuum limits of~$\sqrt{t_{_X}}/w_{0.4}$
and~$w_{_X}/w_{0.4}$ with~$X=0.2$, $0.3$ and~$0.4$ by plotting those quantities
as a function of~$a^2/w_{0.4}^2$ and making linear extrapolation to~$a=0$. Shown
in~Fig.~\ref{fig:wt2} are two examples of such extrapolation for~$X=0.3$. The
resultant values are shown in~Table~\ref{table:ratiotw}, where the statistical
error in the first parenthesis is estimated by $8$-point linear extrapolation,
while the systematic error in the second parenthesis is obtained by the
difference between the $8$-point and $9$-point analyses as mentioned earlier.

\begin{table}[t]
\centering
\begin{tabular}{ccccc}
    \hline \hline
    $\sqrt{t_{0.4}}/w_{0.4}$ & $\sqrt{t_{0.3}}/w_{0.4}$ & $\sqrt{t_{0.2}}/w_{0.4}$ & $w_{0.3}/w_{0.4}$ & $w_{0.2}/w_{0.4}$ \\
    \hline
    1.0164(32)(3) & 0.8785(24)(0) & 0.6952(18)(2) & 0.8968(3)(2) & 0.7665(6)(2) \\
    \hline \hline
\end{tabular}
\caption{Continuum limits of~$\sqrt{t_{_X}}/w_{0.4}$ and~$w_{_X}/w_{0.4}$
for~$X=0.2$, $0.3$ and~$0.4$. The statistical and systematic errors are shown
in the first and second parenthesis, respectively.}
\label{table:ratiotw}
\end{table}

\subsection{Relation to other reference scales}
\label{sec:ratio}

\begin{table}[t]
\centering
\begin{tabular}{ccccc}
    \hline \hline
    $r_c/w_{0.4}$ & $r_0/w_{0.4}$ & $\sqrt{\sigma} w_{0.4}$ &
    $T_c w_{0.4}$  & $w_{0.4}\Lambda_{\overline{\mathrm{MS}}}$ \\
    \hline
  1.328(21)(7) & 2.587(45) & 0.455(8)  & 0.285(5) & 0.233(19) \\
    \hline \hline
\end{tabular}
\caption{Relations of~$w_{0.4}$ with other reference scales in the continuum
limit. For the details of the determination of these values and the estimate
of the errors, see the text.}
\label{table:ratio}
\end{table}

\begin{figure}
\centering
\includegraphics[width=0.6\textwidth,clip]{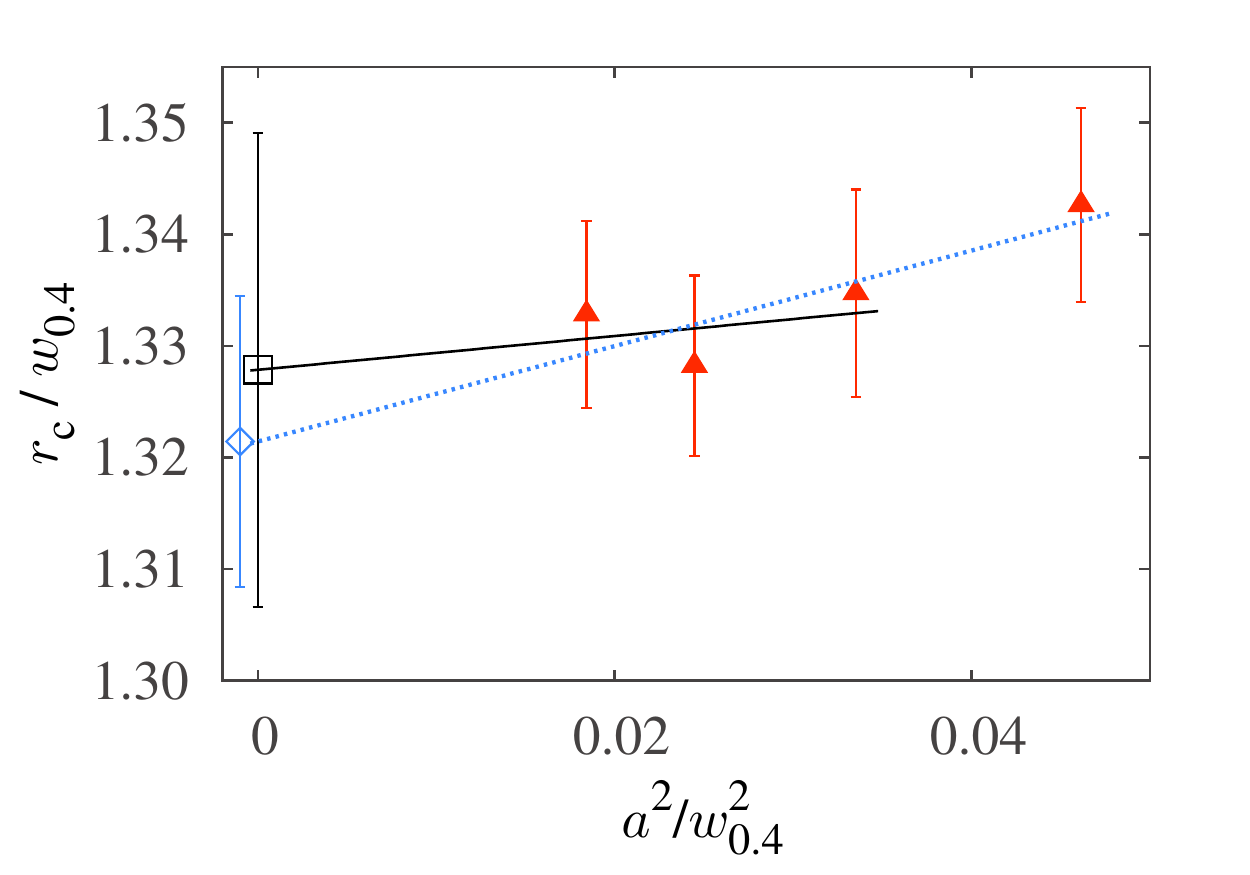}
\caption{Continuum extrapolation of the ratio~$r_c/w_{0.4}$ using $3$-point and
$4$-point linear fits.}
\label{fig:rc}
\end{figure}

We now relate $w_{0.4}$ with other scales used in the literature.
In~Ref.~\cite{Necco:2001xg}, a scale $r_c$ determined from the force~$F(r)$
between heavy quarks as~$r_c^2F(r_c)=0.65$ is introduced and $r_c/a$ was
measured for four $\beta$ values in the range~$6.57\le\beta\le6.92$. Using
Eq.~(\ref{eq:fitfinal}), the result can be converted to the ratio~$r_c/w_{0.4}$,
which are plotted in~Fig.~\ref{fig:rc}. The continuum extrapolation
$r_c/w_{0.4}|_{a\to0}$ obtained from the figure is given in the first column
of~Table \ref{table:ratio}, where the error in the second parenthesis includes
the systematic error from the linear fit and also the uncertainly from the
fitting function in~Eq.~(\ref{eq:fitfinal}).
 
The relation between $w_{0.4}$ and the Sommer scale~$r_0$ defined
by~$r_0^2F(r_0)=1.65$ is also obtained by~$r_c/r_0=0.5133(24)$ given
in~Ref.~\cite{Necco:2001xg}; this is shown in the second column
of~Table \ref{table:ratio}, where the error takes into account all statistical
and systematic ambiguities. The relation between $r_0$ and the string
tension~$\sqrt{\sigma}$ is studied in~Ref.~\cite{Edwards:1997xf} with the
result $r_0\sqrt{\sigma}=1.178$ within~$1\%$ uncertainty. The resulting value
of~$\sqrt{\sigma}w_{0.4}$ is shown in the third column
of~Table~\ref{table:ratio}. We note that the continuum-extrapolated value
of~$\sqrt{t_0}/r_0$ is estimated on coarser lattices
in~Refs.~\cite{Luscher:2010iy,Ce:2014sfa}. These values are consistent with our
results within statistical errors.

In~Table~\ref{table:ratio}, we also show the relations of~$w_{0.4}$ with the
critical temperature of the deconfinement transition~$T_c$ and lambda
parameter~$\Lambda_{\overline{\mathrm{MS}}}$ in the $\overline{\mathrm{MS}}$
scheme, where we used
$T_c/\sqrt{\sigma}=0.625\pm0.003(+0.004)$~\cite{Boyd:1996bx}
and~$r_0\Lambda_{\overline{\mathrm{MS}}}=0.602(48)$~\cite{Capitani:1998mq}. We
note that, in~Ref.~\cite{Boyd:1996bx}, the value of~$\beta$ corresponding to
the critical temperature with~$N_\tau=12$ is obtained as~$\beta_c=6.3384$. This
together with~Eq.~(\ref{eq:fitfinal}) leads to
\begin{equation}
   T_cw_{0.4}=0.2826(3),
\label{eq:Tcw04}
\end{equation}
which is consistent with the value in~Table~\ref{table:ratio}.

We note that $\Lambda_{\overline{\mathrm{MS}}}$ can be also determined by matching
the tadpole improved coupling constant to that in the
${\overline{\mathrm{MS}}}$-scheme~\cite{Gockeler:2005rv}.
In~Appendix~\ref{sec:lambdaMSbar}, we estimated the
ratio~$w_{0.4}\Lambda_{\overline{\mathrm{MS}}}$ by the same analysis
as in~Ref.~\cite{Gockeler:2005rv} using our numerical data on plaquette
and~$w_{0.4}/a$: The result reads
\begin{equation}
   w_{0.4}\Lambda_{\overline{\mathrm{MS}}}=0.2388(5)(13),
\label{eq:w04LambdaMSbar}
\end{equation}
which is consistent with the result in~Table~\ref{table:ratio}.

\subsection{Relation to other parametrizations}
\label{sec:para-ratio}

\begin{figure}[]
\centering
\includegraphics[width=0.6\textwidth,clip]{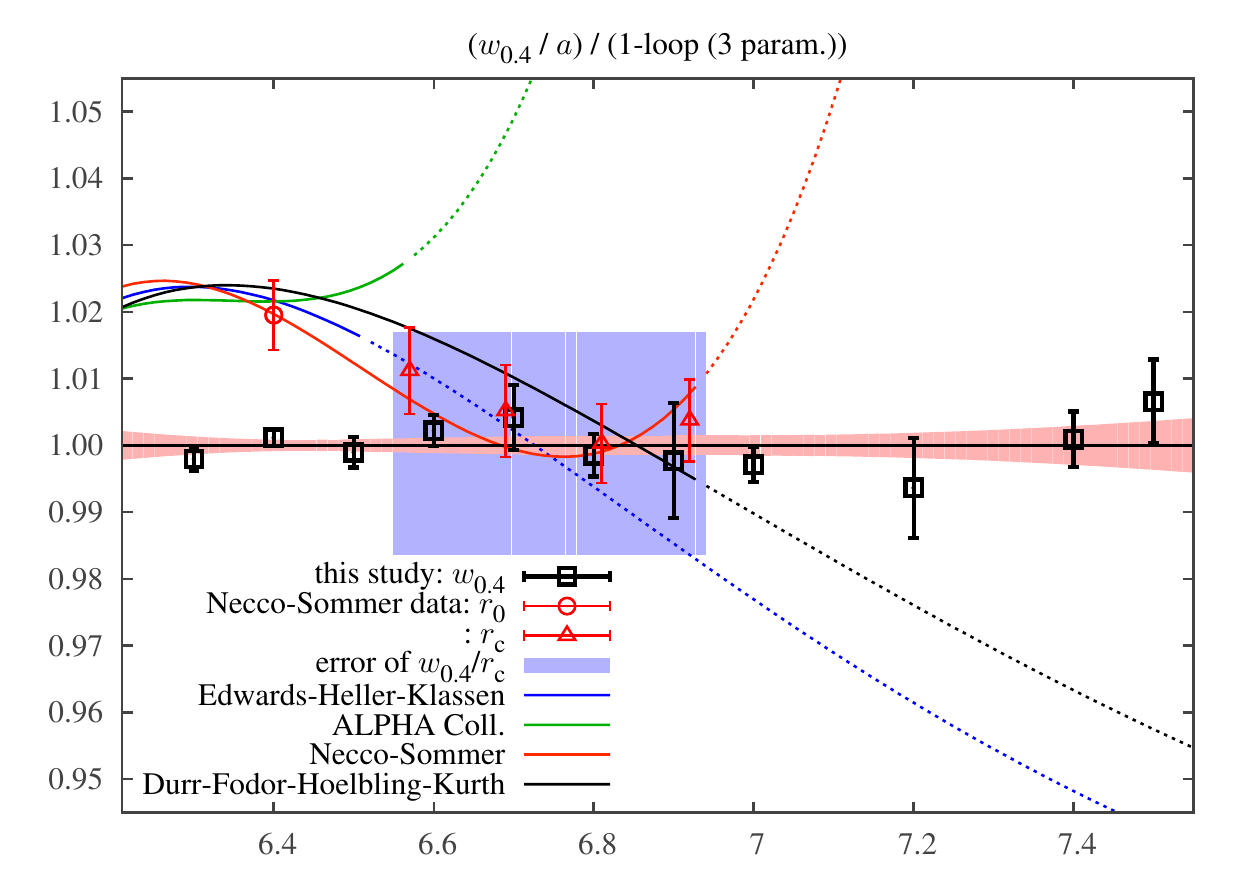}
\caption{Comparison of the fitting function Eq.~(\ref{eq:fitfinal}) with
previous results
in~Refs.~\cite{Edwards:1997xf,Guagnelli:1998ud,Necco:2001xg,Durr:2006ky}. The
numerical results on~$w_{0.4}/a$ measured in the present study and $r_c/a$
and~$r_0/a$ in~Ref.~\cite{Necco:2001xg} are also plotted.
$r_c/a$ and~$r_0/a$ are converted to~$w_{0.4}$-scale using the ratios
in~Table~\ref{table:ratio}.}
\label{fig:comp}
\end{figure}

Let us compare our parametrization Eq.~(\ref{eq:fitfinal}) with those
introduced in previous studies; Edwards, Heller and~Klassen
($5.6\le\beta\le6.5$)~\cite{Edwards:1997xf}, Alpha Collaboration
($5.7\le\beta\le6.57$)~\cite{Guagnelli:1998ud}, Necco and Sommer
($5.7\le\beta\le6.92$)~\cite{Necco:2001xg}, and D\"urr, Fodor, Hoelbling
and~Kurth ($5.7\le\beta\le6.92$)~\cite{Durr:2006ky}. Each parametrization
is based on the data obtained in the range of~$\beta$ given in the parentheses.
In~Refs.~\cite{Guagnelli:1998ud} and~\cite{Necco:2001xg}, fitting functions of
the polynomial form are used, while in~Refs.~\cite{Edwards:1997xf}
and~\cite{Durr:2006ky} the fitting functions motivated by the perturbative
formula are employed.

In~Fig.~\ref{fig:comp}, we show the parametrizations of the above four
references normalized by ours, Eq.~(\ref{eq:fitfinal}). For the conversion
among different reference scales, we have used the ratios
in~Table~\ref{table:ratio}. The error of~$r_c/w_{0.4}$, which dominates the
ambiguity in the relations between $w_{0.4}$ and other scales,
is also shown by the shaded box
in the figure. The figure shows that the parametrizations in the previous
studies agree with ours within this error band in the range of~$\beta$, at which
both fitting functions are reliable. The parametrizations
in~Refs.~\cite{Guagnelli:1998ud} and~\cite{Necco:2001xg} using polynomial
ans\"atze have significant deviation from ours for~$\beta$ outside the
applicable range, while those of the parametrizations
in~Refs.~\cite{Edwards:1997xf} and~\cite{Durr:2006ky} are much milder at the
level of~$4-6\%$ for~$\beta=7.5$.

In~Fig.~\ref{fig:comp}, the numerical results on~$r_0/a$ and~$r_c/a$
in~Ref.~\cite{Necco:2001xg} converted to~$w_{0.4}/a$ using the values
in~Table~\ref{table:ratio} are also presented. The error bars of these points
are the statistical errors in~Ref.~\cite{Necco:2001xg}, and do not include the
ones associated with~$r_c/w_{0.4}$ and~$r_0/w_{0.4}$. The figure indicates that
the $r_0/a$ and $r_c/a$ in~Ref.~\cite{Necco:2001xg} systematically deviate
from our results as~$\beta$ becomes smaller. This may come from the discretization
effect associated with the determination of~$r_0$ on the lattice; 
these different parametrizations do not need to agree, because
different determinations of the reference scales 
can differ by discretization effect.
We also note
that the statistical error in our numerical analysis of~$w_{0.4}/a$ is
significantly smaller than the previous ones for~$6.4\lesssim\beta\lesssim7.0$.

\section{Summary}
\label{sec:summary}
In this paper, we have performed an analysis of the flow time, $t$, dependence 
of~$t^2\langle E(t)\rangle$ and its logarithmic derivative for SU(3)
Yang--Mills theory with~$6.3\le\beta\le7.5$ in large lattice volumes
$N_{\rm s}^4$ ($N_{\rm s}=64$--$128$). The results were utilized to parametrize the $\beta$
dependence of the lattice spacing, Eq.~(\ref{eq:fitfinal}). In our analysis,
the reference scale is chosen to be $w_{0.4}$, which 
is expected to suffer less discretization
error than commonly used $w_{0.3}$ and~$t_{0.3}$. The discretization and finite
volume errors on our results are well suppressed with the present numerical
settings.

\

After the completion of this paper, we found the paper~\cite{Francis:2015lha}
in which $T_c\sqrt{t_{0.3}}=0.2489(14)$ in SU(3) Yang--Mills theory is
obtained. This is consistent with~$T_c\sqrt{t_{0.3}}=0.2483(7)$ obtained from
the values in~Table~\ref{table:ratiotw} and~Eq.~(\ref{eq:Tcw04}).

%
%
%
%

\section*{Acknowledgements}

We would like to thank Etsuko Itou and Tetsuo Hatsuda 
for helpful discussions and comments.
M.~K. thanks H.~Ohno for valuable discussions.
The authors thank the Yukawa Institute for Theoretical Physics,
Kyoto University, where this work was completed
during the YITP-T-14-03 on ``Hadrons and Hadron Interactions in QCD.''
Numerical simulation for this study was carried out on IBM System Blue Gene
Solution at KEK under its Large-Scale Simulation Program (No.~14/15-08). 
This work is supported in part by JSPS KAKENHI Grant Numbers 
23540307, 23540330, 
25287046, 
25800148, 26400272.
T.~I. is supported in part by Strategic Programs for Innovative Research
(SPIRE) Field 5. 

%
%

\appendix
\section{Fit functions}
\label{sec:fit}
In this appendix, we show the four fit functions we employed to parametrize
$w_{0.4}/a$ in terms of~$\beta$:
\begin{enumerate}
\item Polynomials of~$\beta-\beta_0$:
\begin{equation}
   \log\left(\frac{w_{0.4}}{a}\right)_{\mathrm{poly}}(\beta)
   =a_0+a_1(\beta-\beta_0)+a_2(\beta-\beta_0)^2+a_3(\beta-\beta_0)^3.
\label{eq:poly}
\end{equation}

\item One-loop perturbation + polynomials of~$1/\beta$:
\begin{equation}
   \log\left(\frac{w_{0.4}}{a}\right)_{\text{1-loop}}(\beta)
   =\frac{4\pi^2}{33}\beta+c_0+\frac{c_1}{\beta}
   +\frac{c_2}{\beta^2}+\frac{c_3}{\beta^3}+\cdots.
\label{eq:1-loop}
\end{equation}

\item One-loop perturbation + Pad\'e improved polynomials of~$1/\beta$ :
\begin{equation}
  \log \left(\frac{w_{0.4}}{a}\right)_{\text{Pad\'e}}(\beta)
  =\frac{4\pi^2}{33}\beta 
  +d_0\frac{1+d_1/\beta+d_2/\beta^2}{1+e_1/\beta+e_2/\beta^2}.
\label{eq:Pade}
\end{equation}

\item Two-loop perturbation + polynomials of~$1/\beta$:
\begin{equation}
  \log\left(\frac{w_{0.4}}{a}\right)_{\text{2-loop}}(\beta)
  =\frac{4\pi^2}{33}\beta-\frac{51}{121}\ln\beta
  +c'_0+\frac{c'_1}{\beta}+\frac{c'_2}{\beta^2}+\cdots.
\label{eq:2-loop}
\end{equation}

\end{enumerate}
Here, $a_i$, $c_i$, $d_i$, $e_i$ and~$c'_i$ are fitting parameters. The
polynomial form Eq.~(\ref{eq:poly}) is the one used
in~Refs.~\cite{Necco:2001xg} and~\cite{Guagnelli:1998ud}.
In the three-parameter fit with Eqs.~(\ref{eq:poly}), 
(\ref{eq:1-loop}) and (\ref{eq:2-loop}), we set $a_3$, $c_3$ and 
$c'_3$ to zero, respectively.
In the fit with Eq.~(\ref{eq:Pade}), we tried three parameter
fit with $d_2=e_2=0$, four parameter fits with $e_2=0$ and $d_2=0$.
We refer each fit to as [1,1], [2,1] and [1,2], respectively.

\begin{table}
\centering
\begin{tabular}{lc}
    \hline \hline
    fit func. & $\chi^2$/dof \\
    \hline
    polynomial (3 param.) & 1.496 \\
    polynomial (4 param.) & 0.336 \\
    \hline
    1-loop (3 param.) & 0.917 \\
    1-loop (4 param.) & 0.363 \\
    \hline
    $[1,1]$ Pad\'e (3 param.) & 0.377 \\
    $[2,1]$ Pad\'e (4 param.) & 0.424 \\
    $[1,2]$ Pad\'e (4 param.) & 0.424 \\
    \hline \hline
\end{tabular}
\caption{Result of~$\chi^2/{\mathrm{dof}}$ of the fits for~$w_{0.4}/a$ with
various fitting functions Eqs.~(\ref{eq:poly})--(\ref{eq:Pade}).}
\label{table:chi2}
\end{table}

In~Table~\ref{table:chi2} we show $\chi^2/{\mathrm{dof}}$ for each fit
functions. For the polynomial fit, Eq.~(\ref{eq:poly}), we take $\beta_0=7.0$;
the quality of the fit with other choices of~$\beta_0$ hardly changes. In the
text, we employ the three parameter fit with~Eq.~(\ref{eq:1-loop}) to obtain
Eq.~(\ref{eq:fitfinal}), since it provide reasonable
$\chi^2/{\mathrm{dof}}\simeq1$ without over fitting.

\section{Determination of $\Lambda$-parameter}
\label{sec:lambdaMSbar}

In this appendix, we show the derivation of~Eq.(\ref{eq:w04LambdaMSbar}).
In~Ref.~\cite{Gockeler:2005rv}, $r_0\Lambda_{\overline{\mathrm{MS}}}$ is analyzed
with the data of~$r_0/a$ in~Ref.~\cite{Necco:2001xg}. Here we adopt the same
procedure by using the numerical results of~$w_{0.4}/a$ and the average
plaquette obtained in this study. Such an analysis allows us to determine the
ratio $w_{0.4}\Lambda_{\overline{\mathrm{MS}}}$ directly using the accurate data on
fine lattices.

The dimensionless parameter $a\Lambda_{\overline{\mathrm{MS}}}$ can be obtained by
matching the tadpole improved lattice perturbation theory. The boosted coupling
constant~$g_\square$ is defined by
\begin{equation}
   g_\square^2\equiv g_0^2(a)/u_0^4,
\label{}
\end{equation}
where $u_0^4\equiv P=\langle\mathrm{Tr}\ U_\square\rangle/3$.

As for the choice of the renormalization scale and the running coupling
constant, we take the following two methods:
\begin{itemize}
\item Method~I
\begin{equation}
   a\Lambda_{\overline{\mathrm{MS}}}
   =a\mu_\ast F^{\overline{\mathrm{MS}}}(g_{\overline{\mathrm{MS}}}(\mu_\ast))
\label{}
\end{equation}
at the scale 
\begin{equation}
   a\mu_\ast=\exp\left(\frac{t_1^\square}{2b_0}\right),
\label{}
\end{equation}
and 
\begin{equation}
   \frac{1}{g_{\overline{\mathrm{MS}}}^2(\mu_\ast)}
   =\frac{1}{g_\square^2(a)}
   +\left(\frac{b_1}{b_0}t_1^\square-t_2^\square\right)g_\square^2(a)
   +O(g_\square^4).
\label{}
\end{equation}
\item Method~II
\begin{equation}
   a\Lambda_{\overline{\mathrm{MS}}}
   =a\Lambda_\square\exp\left(\frac{t_1^\square}{2b_0}\right),
\label{}
\end{equation}
with 
\begin{equation}
   a\Lambda_\square=F^\square(g_\square(a)).
\label{}
\end{equation}
This scheme corresponds to choosing a scale at
\begin{equation}
   a\mu_=
   =\exp\left(\frac{t_1^\square}{2b_0}\right)
   \frac{F^\square(g_\square(a))}{F^{\overline{\mathrm{MS}}}(g_\square(a)}
\label{}
\end{equation}
in Method~I.
\end{itemize}
In the $3$-loop order, $F^S$ ($S=\square$, $\overline{\mathrm{MS}}$) is
expressed as
\begin{align}
   \frac{\Lambda^S}{M}
   &\equiv F^S(g_S(M))
   =\exp\left(-\frac{1}{2b_0g_S^2}\right)(b_0 g_S^2)^{-\frac{b_1}{2b_0}}
\notag\\
   &\qquad{}\times
   \left(1+\frac{b_1+\sqrt{b_1^2-4b_0b_s^S}}{2b_0}g_S^2\right)^{-p_A^S}
   \left(1+\frac{b_1+\sqrt{b_1^2+4b_0b_s^S}}{2b_0}g_S^2\right)^{-p_B^S},
\label{}
\end{align}
where
\begin{equation}
   p_A^S
   =-\frac{b_1}{4b_0^2}-\frac{b_1^2-2b_0b_2^S}{4b_0^2\sqrt{b_1^2-4b_0b_2^S}},
   \qquad
   p_B^S
   =-\frac{b_1}{4b_0^2}+\frac{b_1^2-2b_0 b_2^S}{4b_0^2\sqrt{b_1^2-4b_0b_2^S}}.
\label{}
\end{equation}
In the $[1,1]$ Pad\'e approximation, it leads
\begin{equation}
   F_{[1,1]}^S(g_S(M))
   =\exp\left(-\frac{1}{2b_0 g_S^2}\right)
   \left[\frac{b_0g_S^2}{1+\left(\frac{b_1}{b_0}-\frac{b_2^S}{b_1}\right)g_S^2}
   \right]^{-\frac{b_1}{2b_0}}.
\label{}
\end{equation}
In SU(3) Yang--Mills theory, the coefficients are given by
\begin{equation}
   b_0=\frac{11}{(4\pi)^2},\qquad b_1=\frac{102}{(4\pi)^4},\qquad
   b_2^{\overline{\mathrm{MS}}}=\frac{1}{(4\pi)^6}\frac{2857}{2},\qquad
   b_2^\square=b_2^{\overline{\mathrm{MS}}}+b_1t_1^\square-b_0t_2^\square,
\label{}
\end{equation}
with
\begin{equation}
   t_1^\square=0.1348680,\qquad t_2^\square=0.0217565.
\label{}
\end{equation}

\begin{figure}
\centering
\includegraphics[width=0.60\textwidth,clip]{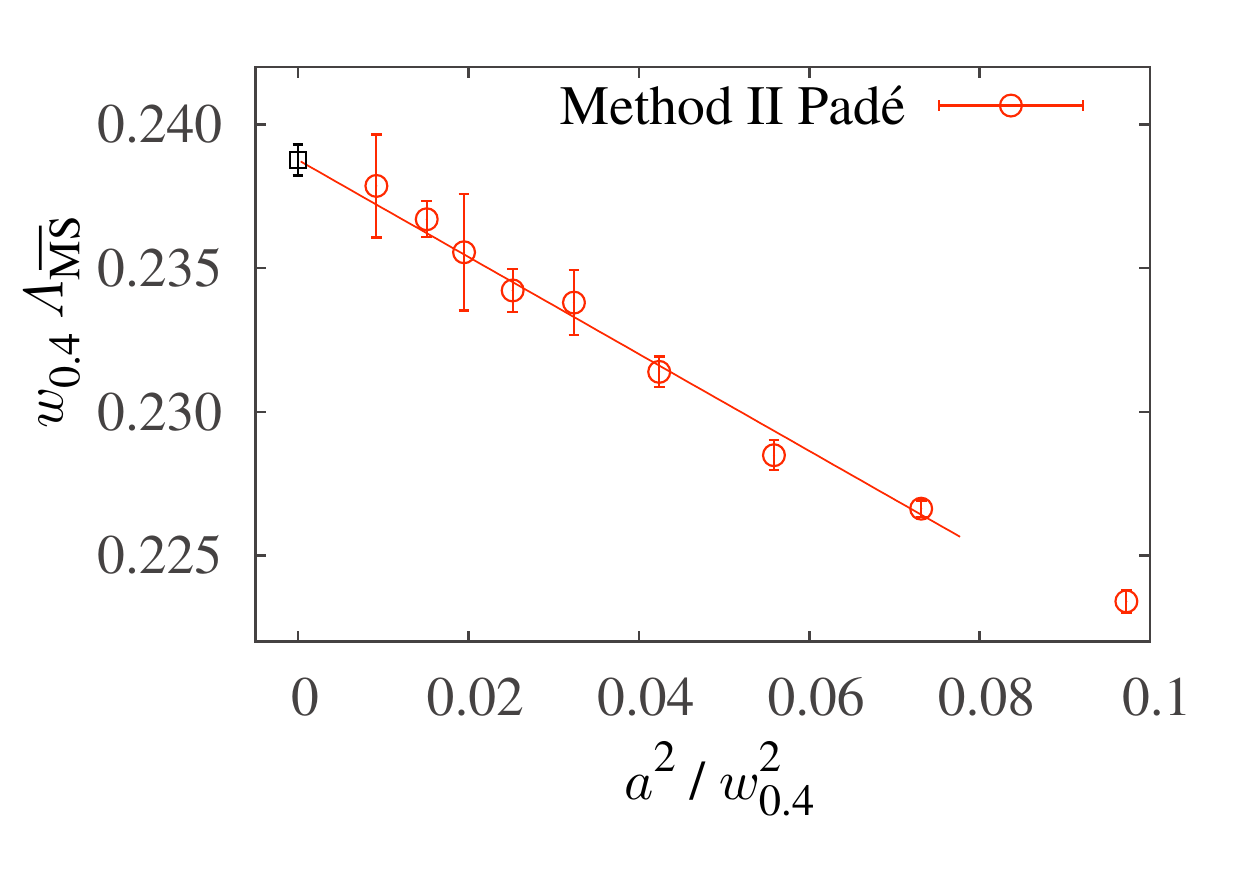}
\caption{Values of~$w_{0.4}\Lambda_{\overline{\mathrm{MS}}}$ estimated by the
Method~II with Pad\'e improvement. The continuum limit is shown
at~$a^2/w_{0.4}^2=0$.}
\label{fig:lambdaMSbarIIPade}
\end{figure}

We adopt Method~II with Pad\'e improvement to estimate the central value
of~$w_{0.4}\Lambda_{\overline{\mathrm{MS}}}$, and use the results of Methods I
and~II without Pad\'e improvement to estimate the systematic error. The results
are summarized in~Table~\ref{tab:lambdaMSbar}: The values in the continuum
limit are obtained by a linear fit as a function of~$a^2/w_{0.4}^2$ without the
coarsest lattice data (see~Fig.~\ref{fig:lambdaMSbarIIPade}). Then we find
\begin{equation}
   w_{0.4}\Lambda_{\overline{\mathrm{MS}}}=0.2388(5)(13)
\label{}
\end{equation}
with the statistical and systematic errors.

\begin{table}
\centering
\begin{tabular}{rrlrlll}
    \hline \hline
    $\beta$ & $N_{\rm s}$ & \multicolumn{1}{c}{plaquette} & \multicolumn{1}{c}{$w_{0.4}/a$} & \multicolumn{3}{c}{$w_{0.4} \Lambda_{\overline{\mathrm{MS}}}$} \\ \cline{5-7}
      & & & & Method I & Method II & Method II Pad\'e \\ 
    \hline
    6.3 & 64 &   0.622 420 85(30)   &  3.208( 7)  &  \textit{0.2248(4)}   
    &  \textit{0.2253(4)}      &   \textit{0.2234(4)} \\
6.4 & 64 &   0.630 632 88(13)   &  3.697( 5)  &  0.2280(3)   &  0.2285(3)      &   0.2266(3) \\
6.5 & 64 &   0.638 361 33(35)   &  4.231(10)  &  0.2298(5)   &  0.2302(5)      &   0.2285(5) \\
6.6 & 64 &   0.645 669 58(12)   &  4.857(11)  &  0.2327(5)   &  0.2331(5)      &   0.2314(5) \\
6.7 & 64 &   0.652 608 39(39)   &  5.558(27)  &  0.2351(11)  &  0.2354(11)     &  0.2338(11) \\
6.8 & 64 &   0.659 215 11(11)   &  6.300(20)  &  0.2354(7)   &  0.2358(8)      &   0.2342(7) \\
6.9 & 64 &   0.665 522 54(33)   &  7.165(62)  &  0.2367(20)  &  0.2370(20)     &  0.2355(20) \\
7.0 & 96 &   0.671 556 729(89)  &  8.137(21)  &  0.2378(6)   &  0.2381(6)      &   0.2367(6) \\
7.2 & 96 &   0.682 891 86(22)   & 10.428(78)  &  0.2389(18)  &  0.2392(18)     &  0.2379(18) \\
\hline
$\infty$ & & 1 & $\infty$ & 0.2399(5) & 0.2401(5) & 0.2388(5)  \\
    \hline \hline
\end{tabular}
\caption{Simulation parameters $\beta$ and~$N_{\rm s}$. The plaquette value,
$w_{0.4}/a$, and~$w_{0.4}\Lambda_{\overline{\mathrm{MS}}}$ using Method I, II and~II
with Pad\'e approximation. The last row corresponds to the values at the
continuum limit obtained from linear extrapolation without using the coarsest
lattice data at~$\beta=6.3$ (the italic number).}
\label{tab:lambdaMSbar}
\end{table}

\end{document}